\documentclass[prd,twocolumn,showpacs,showkeys,preprintnumbers,floatfix,
  nofootinbib]{revtex4-1}
\usepackage{amsfonts} 
\usepackage{amssymb} 
\usepackage{amsmath} 
\usepackage{subfigure} 
\usepackage{graphicx} 
\usepackage{array} 
\usepackage{dcolumn} 
\usepackage{bm} 
\usepackage{latexsym} 
\usepackage{longtable} 
\usepackage{hyperref} 
\usepackage[usenames,dvipsnames]{xcolor}
\usepackage{mathrsfs}
\usepackage{mathtools} 
\usepackage{comment}
\graphicspath{{./figs/}}

\allowdisplaybreaks

\newcommand{\Tr}{\textrm{Tr}}

\newcommand{\mc}[1]{\mathcal{#1}}
\newcommand{\mr}[1]{\mathrm{#1}}
\renewcommand{\mathcal}[1]{\mathscr{#1}}
\newcommand{\val}[1]{\ensuremath{\overline{#1}}}
\newcommand{\sea}[1]{\ensuremath{\underline{#1}}}
\newcommand{\bra}[1]{\langle#1|}
\newcommand{\ket}[1]{|#1\rangle}

\definecolor{HLBlue}{HTML}{6599FF}
\definecolor{HLOrange}{HTML}{FF6600}
\definecolor{dkgreen}{HTML}{088A08}

\newcommand{\ohf}{\ensuremath{\tfrac{1}{2}}}

\newcommand{\wlee}[1]{\textcolor{blue}{\textbf{#1}}} 
\newcommand{\jab}[1]{{#1}} 
\newcommand{\org}[1]{} 

\begin{document}

\title{
Masses and decay constants of pions and kaons in mixed-action staggered chiral perturbation theory
}
\author{Jon~A.~Bailey}
\email[Email: ]{jonbailey@snu.ac.kr}
%
\affiliation{
 Lattice Gauge Theory Research Center, FPRD, and CTP \\
 Department of Physics and Astronomy,
 Seoul National University,
 Seoul, 151-742, South Korea
}
\author{Hyung-Jin~Kim}
%
%
\affiliation{
 Physics Department,
 Brookhaven National Laboratory,
 Upton, NY~~11973, USA
}
\author{Jongjeong~Kim}
%
%
\affiliation{
 Lattice Gauge Theory Research Center, FPRD, and CTP \\
 Department of Physics and Astronomy,
 Seoul National University,
 Seoul, 151-742, South Korea
}
\author{Weonjong~Lee}
\email[Email: ]{wlee@snu.ac.kr}
%
%
\affiliation{
 Lattice Gauge Theory Research Center, FPRD, and CTP \\
 Department of Physics and Astronomy,
 Seoul National University,
 Seoul, 151-742, South Korea
}
\author{Boram Yoon}
%
%
\affiliation{
Los Alamos National Laboratory, Theoretical Division T-2, MS B283
Los Alamos, NM~~87545, USA
}
\collaboration{SWME Collaboration}
\date{\today}
\begin{abstract}
Lattice QCD calculations with different staggered valence and sea quarks can be used to improve determinations of quark masses, Gasser-Leutwyler couplings, and other parameters relevant to phenomenology. We calculate the masses and decay constants of flavored pions and kaons through next-to-leading order in staggered-valence, staggered-sea mixed-action chiral perturbation theory. We present the results in the valence-valence and valence-sea sectors, for all tastes. As in unmixed theories, the taste-pseudoscalar, valence-valence mesons are exact Goldstone bosons in the chiral limit, at non-zero lattice spacing.  The results reduce correctly when the valence and sea quark actions are identical, connect smoothly to the continuum limit, and provide a way to control light quark and gluon discretization errors in lattice calculations performed with different staggered actions for the valence and sea quarks.
\end{abstract}
\maketitle
%
\section{\label{sec:intro}Introduction}
 
The quark masses and CKM matrix elements are fundamental parameters
of the Standard Model.  To understand their values in terms of the
underlying physics and probe the limits of the Standard Model, they
must be extracted from experiment with greater precision.  In
addition, the low-energy couplings (LECs) of chiral perturbation
theory (ChPT) parametrize the strong interactions at energies small
compared to the scale of chiral symmetry breaking \cite{
  Weinberg:1978kz, Gasser:1983yg, Gasser:1984gg}.  Improving knowledge
of the Standard Model and chiral effective theory parameters requires
improved calculations of strong force contributions to the relevant
hadronic matrix elements.

Mixed-action lattice QCD calculations can be used to calculate
hadronic matrix elements while exploiting the advantages of different
discretizations of the fermion action.  For example, fermions with
more desirable features for a specific physics purpose may be used for
the valence quarks, while fermions more adequate for massive production
may be used for the sea quarks, to include the effects of vacuum
polarization.  The construction of chiral effective theories for
lattice QCD incorporates discretization effects, thereby relating the
chiral and continuum extrapolations and improving control of the
continuum limit~\cite{Lee:1999zxa,Rupak:2002sm}.

Staggered ChPT (SChPT) was developed to analyze results of lattice calculations
with staggered fermions~\cite{Lee:1999zxa,Aubin:2003mg}; it has been used
extensively to control extrapolations to physical light-quark masses and to
remove dominant light-quark and gluon discretization
errors~\cite{Bazavov:2009bb}.  Mixed-action ChPT was developed for lattice
calculations performed with Ginsparg-Wilson valence quarks and Wilson sea
quarks~\cite{Bar:2002nr,Bar:2003mh}.  The formalism for staggered sea quarks
and Ginsparg-Wilson valence quarks was developed in Ref.~\cite{Bar:2005tu}.
Mixed-action ChPT for differently improved staggered fermions was introduced
for calculations of the $K^0-\overline{K^0}$ mixing bag parameters entering
$\varepsilon_K$ in and beyond the Standard Model \cite{ Bae:2010ki,
Bailey:2012wb,Bae:2013tca} and the $K\to\pi\ell\nu$ vector form factor
\cite{Bazavov:2012cd}.

We have calculated the pion and kaon masses and axial-current decay constants
in all taste representations at next-to-leading order (NLO) in mixed-action
SChPT.  The results generalize those of
Refs.~\cite{Aubin:2003mg,Aubin:2003uc,SWME:2011aa,Bailey:2012jy} to the
mixed-action case; the results could be used to improve determinations of LECs
poorly determined by existing analyses and to improve determinations of
light-quark masses, the Gasser-Leutwyler couplings, and the pion and kaon decay
constants.

In Sec.~\ref{sec:maschpt} we review the formulation of mixed-action SChPT.
Results for the masses are presented in Sec.~\ref{sec:mass}, and for the decay
constants, in Sec.~\ref{sec:decay}.  In Sec.~\ref{sec:sum}, we conclude.
%
\section{\label{sec:maschpt}Mixed-action staggered chiral perturbation theory}
As for ordinary, unmixed SChPT, the theory is constructed in two steps.  First
one builds the Symanzik effective continuum theory (SET) for the lattice
theory.  Then one maps the operators of the SET into those of
ChPT~\cite{Lee:1999zxa,Aubin:2003mg,Bae:2010ki,CB:notes}.  
\subsection{\label{subsec:set}Symanzik effective theory}
Through NLO the SET may be written
\begin{align}
S_\mr{eff} &= S_\mr{QCD} + a^2 S_6 + \dots\,,
\end{align}
where $S_\mr{QCD}$ has the form of the QCD action, but possesses taste degrees
of freedom and respects the continuum taste SU(4) symmetry.  To account for
differences in the masses of valence and sea quarks in lattice calculations,
the SET can be formulated with bosonic ghost quarks and fermionic valence and
sea quarks~\cite{Bernard:1993sv}.  We use the replica method
\cite{Damgaard:2000gh} and so include in the action only (fermionic)
valence and sea quarks.

The operators in $S_6$ have mass-dimension six, and they break the continuum
symmetries to those of the mixed-action lattice theory.  In valence and sea
sectors, these symmetries are identical to those in the unmixed
case~\cite{Lee:1999zxa,Aubin:2003mg}, but now there are no symmetries rotating
valence and sea quarks together~\cite{Bae:2010ki,CB:notes}.  
As in the unmixed case, only a subset of the operators in $S_6$ contribute to
the leading-order (LO) chiral Lagrangian, and they are four-fermion operators
respecting the remnant taste symmetry $\Gamma_4\rtimes\,$SO(4) $\subset$ SU(4).
They can be obtained from those of the unmixed SET by introducing projection
operators $P_{v,\sigma}$ onto the valence and sea sectors in the
$\Gamma_4\rtimes\,$SO(4)-respecting operators of the unmixed theory and
allowing the LECs in the valence and sea sectors to take different
values~\cite{CB:notes}. 
Generically,
\begin{align}
c\,&\bar{\psi}(\gamma_s\otimes\xi_t)\psi\,\bar{\psi}(\gamma_s\otimes\xi_t)\psi\label{eq:genop} \\
\longrightarrow\,c_{vv}\,&\bar{\psi}(\gamma_s\otimes\xi_t)P_v\psi\,\bar{\psi}(\gamma_s\otimes\xi_t)P_v\psi + (v\rightarrow\sigma) \nonumber\\
+\,2c_{v\sigma}\,&\bar{\psi}(\gamma_s\otimes\xi_t)P_v\psi\,\bar{\psi}(\gamma_s\otimes\xi_t)P_\sigma\psi\,, \nonumber
\end{align}
where $\gamma_s$ ($\xi_t$) is a spin (taste) matrix, and the quark spinors
$\psi$ carry flavor indices taking on values in the valence and sea sectors.
In Eq.~\eqref{eq:genop}, the flavor indices are contracted within each
bilinear.  For the action of the projection operators on the spinors,
we may write
\begin{align}
(P_v\psi)_i &= \psi_i & (P_\sigma\psi)_i &= 0 & \text{for $i\in v$}\,,\\
(P_v\psi)_i &= 0 & (P_\sigma\psi)_i &= \psi_i & \text{for $i\in \sigma$}\,.\nonumber
\end{align}
In the unmixed case, $c_{vv}=c_{\sigma\sigma}=c_{v\sigma}=c$, and we recover
the operators of the unmixed theory.

\subsection{\label{subsec:lo-chlag}Leading order chiral Lagrangian}
Mapping the SET operators into the chiral theory at LO, we may write~\cite{CB:notes}
\begin{align}
\mc{L}_\mr{LO} &= \frac{f^2}{8} \mr{Tr}(\partial_{\mu}\Sigma \partial_{\mu}\Sigma^{\dagger}) -
\frac{1}{4}\mu f^2 \mr{Tr}(M\Sigma+M\Sigma^{\dagger})\label{eq:LOlag}\\
&+ \frac{2m_0^2}{3}[\mr{Tr}(\phi_I)]^2
+ a^2 \mc{V}\,.\nonumber
\end{align}
%
%
The first three terms are identical to the kinetic energy, mass, and anomaly
operators of the unmixed theory, respectively; the normalization of the anomaly
term is arbitrary, but natural in SU(3) SChPT, for which the mass of the
taste-singlet $\eta^\prime$ approaches $m_0$ as
$m_0\rightarrow\infty$~\cite{Aubin:2003mg}.
\jab{As in the unmixed theory, the field $\Sigma = e^{i\phi/f}$, where
\begin{align}
\phi &= \sum_{a=1}^{16} \phi^a \otimes T^a\,,\\
T^a &\in \{\xi_5,\ i\xi_{\mu5},\ i\xi_{\mu\nu}(\mu < \nu),\ \xi_{\mu},\ \xi_I\}\,.
\end{align}
The field $\phi$ is a matrix in flavor-taste space, and the Hermitian, $4\times
4$ generators of (taste) U(4) $T^a$ are defined in terms of the taste matrices
$\xi_\mu$, which generate the Clifford algebra; $\xi_{\mu 5} \equiv
\xi_{\mu}\xi_5$, $\xi_{\mu\nu} \equiv \xi_\mu \xi_\nu$, and $\xi_I \equiv I$,
the identity in taste space.}

To construct the potential $\mc{V}$, the projection operators are
conveniently included in spurions.  The result can be written
\begin{align}
\mc{V} = \mc{U} + \mc{U}^\prime - C_\mr{mix}\mr{Tr}(\tau_3\Sigma\tau_3\Sigma^\dagger),\label{eq:pot}
\end{align}
where the last term is a taste-singlet potential new in the mixed-action
theory, with $\tau_3\equiv P_\sigma - P_v$.  It arises from four-quark
operators in which \org{$\xi_t=I$, the identity in taste
space}\jab{$\xi_t=\xi_I$}; such operators map to constants in the unmixed case.
In the mixed-action theory, they yield nontrivial chiral operators because the
projection operators $P\neq 1$ are included in the taste
spurions~\cite{CB:notes}.  In the appendix we present a derivation of the last
term in Eq.~\eqref{eq:pot}.
%
%
The potentials $\mc{U}$ and $\mc{U}^\prime$ contain single- and
double-trace operators, respectively, that are direct generalizations of those
in unmixed SChPT.  
The operators in $\mc{U}^{(\prime)}$ have independent LECs
for the valence-valence, sea-sea, and valence-sea sectors.  We write
\begin{align}
\mc{U} &= \mc{U}_{vv}+\mc{U}_{\sigma\sigma}+\mc{U}_{v\sigma}\,,\\
\mc{U^\prime} &= \mc{U^\prime}_{vv}+\mc{U^\prime}_{\sigma\sigma}+\mc{U^\prime}_{v\sigma}\,,
\end{align}
where
\begin{align}
-\mc{U}_{vv} &= 
C^{vv}_1\Tr(\xi_5P_v\Sigma\xi_5P_v\Sigma^{\dagger})\\
&+ C^{vv}_6\ \sum_{\mu<\nu} \Tr(\xi_{\mu\nu}P_v\Sigma \xi_{\nu\mu}P_v\Sigma^{\dagger}) \nonumber\\
&+ \frac{C^{vv}_3}{2} \jab{\sum_{\nu}} [ \Tr(\xi_{\nu}P_v\Sigma \xi_{\nu}P_v\Sigma) + p.c.] \nonumber\\
&+ \frac{C^{vv}_4}{2} \jab{\sum_{\nu}} [ \Tr(\xi_{\nu 5}P_v\Sigma \xi_{5\nu}P_v\Sigma) + p.c.] \,,\nonumber
\end{align}
\begin{align}
-\mc{U}_{\sigma\sigma} &= 
C^{\sigma\sigma}_1\Tr(\xi_5P_\sigma\Sigma\xi_5P_\sigma\Sigma^{\dagger})\\
&+ C^{\sigma\sigma}_6\ \sum_{\mu<\nu} \Tr(\xi_{\mu\nu}P_\sigma\Sigma \xi_{\nu\mu}P_\sigma\Sigma^{\dagger}) \nonumber\\
&+ \frac{C^{\sigma\sigma}_3}{2} \jab{\sum_{\nu}} [ \Tr(\xi_{\nu}P_\sigma\Sigma \xi_{\nu}P_\sigma\Sigma) + p.c.] \nonumber\\
&+ \frac{C^{\sigma\sigma}_4}{2} \jab{\sum_{\nu}} [ \Tr(\xi_{\nu 5}P_\sigma\Sigma \xi_{5\nu}P_\sigma\Sigma) + p.c.] \,,\nonumber
\end{align}
\begin{align}
-\mc{U}_{v\sigma} &= 
C^{v\sigma}_1[\Tr(\xi_5P_v\Sigma\xi_5P_\sigma\Sigma^{\dagger}) + p.c.]\\
&+ C^{v\sigma}_6\ \sum_{\mu<\nu}[ \Tr(\xi_{\mu\nu}P_v\Sigma \xi_{\nu\mu}P_\sigma\Sigma^{\dagger}) + p.c.] \nonumber\\
&+ C^{v\sigma}_3 \jab{\sum_{\nu}} [ \Tr(\xi_{\nu}P_v\Sigma \xi_{\nu}P_\sigma\Sigma) + p.c.]\nonumber\\
&+ C^{v\sigma}_4 \jab{\sum_{\nu}} [ \Tr(\xi_{\nu 5}P_v\Sigma \xi_{5\nu}P_\sigma\Sigma) + p.c.]\,, \nonumber
\end{align}
\begin{align}
-\mc{U^\prime}_{vv} &=
\frac{C^{vv}_{2V}}{4} \jab{\sum_{\nu}}[\Tr(\xi_{\nu}P_v\Sigma)\Tr(\xi_{\nu}P_v\Sigma) + p.c.]\\
&+ \frac{C^{vv}_{2A}}{4} \jab{\sum_{\nu}} [ \Tr(\xi_{\nu5}P_v\Sigma)\Tr(\xi_{5\nu}P_v\Sigma) + p.c.] \nonumber\\
&+ \frac{C^{vv}_{5V}}{2} \jab{\sum_{\nu}} [ \Tr(\xi_{\nu}P_v\Sigma)
 \Tr(\xi_{\nu}P_v\Sigma^{\dagger})] \nonumber\\
&+ \frac{C^{vv}_{5A}}{2} \jab{\sum_{\nu}} [ \Tr(\xi_{\nu5}P_v\Sigma)
 \Tr(\xi_{5\nu}P_v\Sigma^{\dagger}) ]\,,\nonumber
\end{align}
\begin{align}
-\mc{U^\prime}_{\sigma\sigma} &=
\frac{C^{\sigma\sigma}_{2V}}{4} \jab{\sum_{\nu}}[\Tr(\xi_{\nu}P_\sigma\Sigma)\Tr(\xi_{\nu}P_\sigma\Sigma) + p.c.]\\
&+ \frac{C^{\sigma\sigma}_{2A}}{4} \jab{\sum_{\nu}} [ \Tr(\xi_{\nu5}P_\sigma\Sigma)\Tr(\xi_{5\nu}P_\sigma\Sigma) + p.c.]\nonumber \\
&+ \frac{C^{\sigma\sigma}_{5V}}{2} \jab{\sum_{\nu}} [ \Tr(\xi_{\nu}P_\sigma\Sigma)
 \Tr(\xi_{\nu}P_\sigma\Sigma^{\dagger})]\nonumber\\
&+ \frac{C^{\sigma\sigma}_{5A}}{2} \jab{\sum_{\nu}} [ \Tr(\xi_{\nu5}P_\sigma\Sigma)
 \Tr(\xi_{5\nu}P_\sigma\Sigma^{\dagger}) ]\,,\nonumber
\end{align}
\begin{align}
-\mc{U^\prime}_{v\sigma} &=
\frac{C^{v\sigma}_{2V}}{2} \jab{\sum_{\nu}}[\Tr(\xi_{\nu}P_v\Sigma)\Tr(\xi_{\nu}P_\sigma\Sigma) + p.c.]\\
&+ \frac{C^{v\sigma}_{2A}}{2} \jab{\sum_{\nu}} [ \Tr(\xi_{\nu5}P_v\Sigma)\Tr(\xi_{5\nu}P_\sigma\Sigma) + p.c.] \nonumber\\
&+ \frac{C^{v\sigma}_{5V}}{2} \jab{\sum_{\nu}} [ \Tr(\xi_{\nu}P_v\Sigma)
 \Tr(\xi_{\nu}P_\sigma\Sigma^{\dagger}) + p.c.]\nonumber\\
&+ \frac{C^{v\sigma}_{5A}}{2} \jab{\sum_{\nu}} [ \Tr(\xi_{\nu5}P_v\Sigma)
 \Tr(\xi_{5\nu}P_\sigma\Sigma^{\dagger}) + p.c.]\,,\nonumber
\end{align}
where $p.c.$ indicates the parity conjugate.  In the unmixed case,
$C_\mr{mix}=0$, $C^{vv}=C^{\sigma\sigma}=C^{v\sigma}=C$, and the potential
$\mc{V}$ reduces to that of ordinary SChPT.  Restricting attention to two-point
correlators of sea-sea particles yields results of the unmixed theory, as
expected~\cite{Bernard:1993sv}.
\subsection{\label{subsec:tree}Tree-level masses and propagators}
As in the unmixed theory, the potential $\mc{V}$ contributes to the tree-level
masses of the pions and kaons, which fall into irreducible representations
(irreps) of $\Gamma_4\rtimes\,$SO(4). 
For a taste $t$ pseudo-Goldstone boson (PGB) $\phi^t_{xy}$ composed of quarks
with flavors $x,y,\ x\neq y$,
%
%
%
\begin{align}
m_{xy,\,t}^2 &= \mu (m_x + m_y) + a^2 \Delta_F^{xy}\,,\label{eq:tree}\\
           t &\in F\in\{P,A,T,V,I\} \,,\nonumber
\end{align}
where $F$ labels the taste $\Gamma_4\rtimes\,$SO(4) irrep (pseudoscalar, axial,
tensor, vector, or scalar).
The notation here matches that in our recent papers \cite{
Bailey:2012jy, SWME:2011aa} on taste non-Goldstone pions and kaons in
  ordinary SChPT.
It is also the basis for the notation in the sections below.
The mass splitting $\Delta_F^{xy}$ depends on the LEC of the taste-singlet
potential ($C_\mr{mix}$), the LECs in the single-trace potential ($\mc{U}$),
and the sector (valence or sea) of the quark flavors $x$ and $y$.
Expanding the LO Lagrangian through $\mc{O}(\phi^2)$, we have
\begin{align}
\Delta^{vv}_F &= \frac{8}{f^2}\sum_{b\neq I}\,C^{vv}_b\,(1-\theta^{tb}\theta^{b5})\,,\label{eq:splt-vv}\\
\Delta^{\sigma\sigma}_F &= \frac{8}{f^2}\sum_{b\neq I}\,C^{\sigma\sigma}_b\,(1-\theta^{tb}\theta^{b5})\,,
\label{eq:splt-ss}\\
\Delta^{v\sigma}_F &=\frac{16C_\mr{mix}}{f^2}+\frac{8}{f^2}\sum_{b\neq I}\left[\tfrac{1}{2}(C^{vv}_b+C^{\sigma\sigma}_b)-C^{v\sigma}_b\theta^{tb}\theta^{b5}\right]\,,
\label{eq:splt-vs}
\end{align}
where the splitting is $\Delta_F^{vv}$ if both quarks are valence quarks
($xy\in vv$), $\Delta_F^{\sigma\sigma}$ if both quarks are sea quarks ($xy\in
\sigma\sigma$), and $\Delta_F^{v\sigma}$ otherwise.  
The sub(super)script $b$ and taste $t$ are indices labeling the generators of
the fundamental irrep of U(4).
The numerical constant $\theta^{tb}=+1$ if the generators for $t$ and $b$
commute and $-1$ if they anti-commute.
The LEC $C_b = C_1$, $C_6$, $C_3$, or $C_4$ if $b$ labels a generator
corresponding to the $P$, $T$, $V$, or $A$ irrep of $\Gamma_4\rtimes\,$SO(4),
respectively.  

The residual chiral symmetry in the valence-valence sector, as for the unmixed
theory, implies $F=P$ particles are Goldstone bosons for $a\neq 0,\ m_q = 0$,
and therefore $\Delta_P^{vv}=0$.  The same is not true for the taste
pseudoscalar, valence-sea PGBs, and generically, $\Delta_P^{v\sigma}\neq 0$. 

In the flavor-neutral sector, $x = y$, the PGBs mix in the taste singlet,
vector, and axial irreps.  The Lagrangian mixing terms (hairpins) are
\begin{align}
&\tfrac{1}{2}\,\delta_I^{ij}\,\phi^I_{ii}\phi^I_{jj} + \tfrac{1}{2}\,\delta_V^{vv}\,\phi^\mu_{\val{i}\val{i}}\phi^\mu_{\val{j}\val{j}} + \tfrac{1}{2}\,\delta_V^{\sigma\sigma}\,\phi^\mu_{\sea{i}\sea{i}}\phi^\mu_{\sea{j}\sea{j}} + \delta_V^{v\sigma}\,\phi^\mu_{\val{i}\val{i}}\phi^\mu_{\sea{j}\sea{j}} \\
&+ (V\rightarrow A,\ \mu \rightarrow \mu5)\,,\nonumber \label{eq:hair}
\end{align}
where $i,j$ are flavor indices; $\mu$ ($\mu5$) is a taste index in the vector (axial) irrep; and we use an overbar (underbar) to restrict summation to the valence (sea) sector.  
The $\delta_I^{ij}$-term is the anomaly term; $\delta_I^{ij}\equiv
4m_0^2/3$.  In continuum ChPT, taking $m_0\rightarrow\infty$ at the end of the
calculation decouples the $\eta^\prime$~\cite{Sharpe:2001fh}.  In SChPT, taking
$m_0\rightarrow\infty$ decouples the $\eta^\prime_I$.  The flavor-singlets in
other taste irreps are PGBs and do not decouple~\cite{Aubin:2003mg}.  
The $\delta_{V,\,A}^{vv,\sigma\sigma,v\sigma}$-terms are lattice artifacts from
the double-trace potential $a^2\mathcal{U}^\prime$, and the couplings
$\delta_{V,\,A}^{vv,\sigma\sigma,v\sigma}$ depend linearly on its LECs,
\begin{align}
\delta^{vv}_V &= \frac{16a^2}{f^2}(C^{vv}_{2V}-C^{vv}_{5V}) & \delta^{vv}_A &= \frac{16a^2}{f^2}(C^{vv}_{2A}-C^{vv}_{5A}) \\
\delta^{\sigma\sigma}_V &= \frac{16a^2}{f^2}(C^{\sigma\sigma}_{2V}-C^{\sigma\sigma}_{5V}) & \delta^{\sigma\sigma}_A &= \frac{16a^2}{f^2}(C^{\sigma\sigma}_{2A}-C^{\sigma\sigma}_{5A})\\
\delta^{v\sigma}_V &= \frac{16a^2}{f^2}(C^{v\sigma}_{2V}-C^{v\sigma}_{5V}) & \delta^{v\sigma}_A &= \frac{16a^2}{f^2}(C^{v\sigma}_{2A}-C^{v\sigma}_{5A})\,.
\end{align}

Although the mass splittings and hairpin couplings are different in the three
sectors, the tree-level propagator can be written in the same form as in the
unmixed case.  We have ($k,l$ are flavor indices)
\begin{align}
G^{tb}_{ij,\,kl}(p^2)=\delta^{tb}\left(\frac{\delta_{il}\delta_{jk}}{p^2+m_{ij,\,t}^2}+\delta_{ij}\delta_{kl}\,D^t_{il}\right),
\label{treeprop}
\end{align}
where the disconnected propagators vanish (by definition) in the pseudoscalar and tensor irreps, and for the singlet, vector, and axial irreps,
\begin{align}
D^t_{ij} &\equiv \frac{-1}{I_t J_t}
\frac{\delta^{ij}_F}{1+\delta^{\sigma\sigma}_F \sigma_t}\quad\text{for $ij\notin vv$}\,,\label{eq:discDnotvv}\\
D^t_{ij} &\equiv \frac{-1}{I_t J_t} 
\left(\frac{(\delta^{v\sigma}_F)^2/\delta^{\sigma\sigma}_F}{1+\delta^{\sigma\sigma}_F\sigma_t}+
\delta^{vv}_F-(\delta^{v\sigma}_F)^2/
\delta^{\sigma\sigma}_F\right)\label{eq:discDvv}\\
 &\text{for $ij\in vv$}\,,\nonumber
\end{align}
where $I_t \equiv p^2+m_{ii,\,t}^2$, $J_t \equiv
p^2+ m_{jj,\,t}^2$, and we use the replica method to quench the valence
quarks~\cite{Damgaard:2000gh} and root the sea quarks~\cite{Aubin:2003mg}, so
that
%
%
\begin{align}
\sigma_t &\equiv \sum_{{i}}\frac{1}{p^2 + m_{{ii},\,t}^2} 
               \rightarrow \tfrac{1}{4}\sum_{\sea{i}^\prime}\frac{1}{p^2 + m_{\sea{i}^\prime\sea{i}^\prime,\,t}^2}\,.
\end{align}
The index $\sea{i}^\prime$ is summed over the physical sea quark flavors.  
As for the continuum, partially quenched case~\cite{Sharpe:2000bc}, the factors
arising from iterating sea quark loops can be reduced to a form convenient for
doing loop integrations.
For three nondegenerate, physical sea quarks $u$, $d$, $s$, we have
\begin{align}
\frac{1}{1+\delta^{\sigma\sigma}_F{\sigma_t}}&=\frac{(p^2+m_{uu,\,t}^2)(p^2+m_{dd,\,t}^2)(p^2+m_{ss,\,t}^2)}{\bigl(p^2+m_{\pi^0_t}^2\bigr)\bigl(p^2+m_{\eta_t}^2\bigr)\bigl(p^2+m_{\eta^\prime_t}^2\bigr)}\,,\label{eq:detrat}
\end{align}
%
%
where $m_{\pi^0_t}^2$, $m_{\eta_t}^2$, and $m_{\eta^\prime_t}^2$ are the eigenvalues of the matrices (for tastes $F = I,V,A$)
\begin{align}
        \begin{pmatrix}
        m_{uu,\,t}^2 + \delta^{\sigma\sigma}_F/4 & \delta^{\sigma\sigma}_F/4 & \delta^{\sigma\sigma}_F/4 \\
        \delta^{\sigma\sigma}_F/4 & m_{dd,\,t}^2 + \delta^{\sigma\sigma}_F/4 & \delta^{\sigma\sigma}_F/4 \\
        \delta^{\sigma\sigma}_F/4 & \delta^{\sigma\sigma}_F/4 & m_{ss,\,t}^2 + \delta^{\sigma\sigma}_F/4
        \end{pmatrix}\,.
\end{align}
%
%
In the disconnected propagator $D^t_{ij}$, an additional piece appears
in the valence-valence sector (Eq.~\eqref{eq:discDvv}).  As noted in
Refs.~\cite{Bae:2010ki,CB:notes}, this piece has the form of a
quenched disconnected propagator, for which ${\sigma_t}=0$, and the
assumption of factorization leads us to expect its suppression\jab{;
  by comparing results of analyses with SU(2) and SU(3) mixed-action
  SChPT, the authors of Ref.~\cite{Bae:2010ki} showed the associated
  contributions to $B_K$ were negligible compared to other
  uncertainties.}  In the unmixed case, the mass splittings and
hairpin couplings in the valence and sea sectors are degenerate, and
the propagator reduces.

\begin{figure}[htbp!]
\subfigure[]{\includegraphics[width=0.17\textwidth]{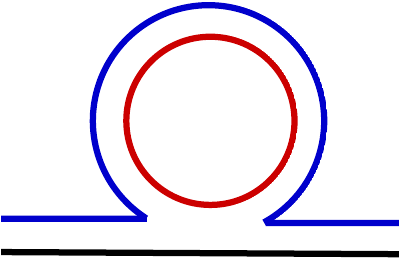}}
\subfigure[]{\includegraphics[width=0.17\textwidth]{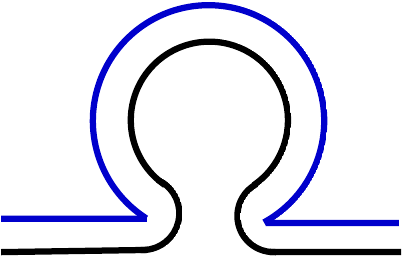}}
\subfigure[]{\includegraphics[width=0.17\textwidth]{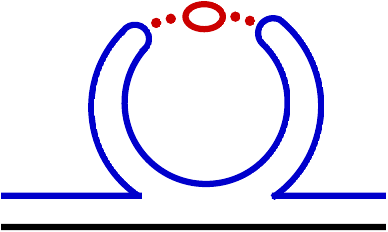}}
\subfigure[]{\includegraphics[width=0.17\textwidth]{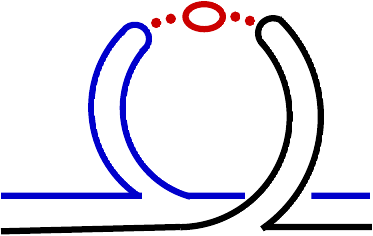}}
\subfigure[]{\includegraphics[width=0.17\textwidth]{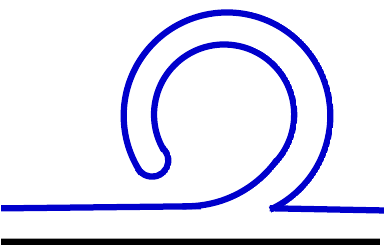}}
\subfigure[]{\includegraphics[width=0.17\textwidth]{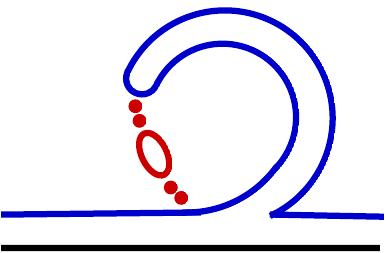}}
\subfigure[]{\includegraphics[width=0.2\textwidth]{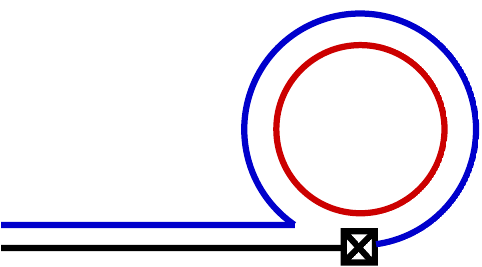}}
\subfigure[]{\includegraphics[width=0.2\textwidth]{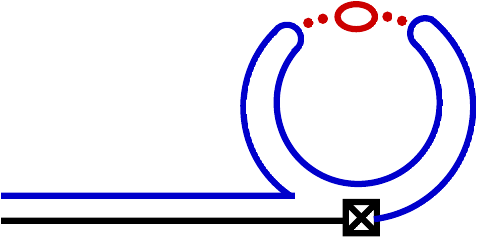}}
\subfigure[]{\includegraphics[width=0.2\textwidth]{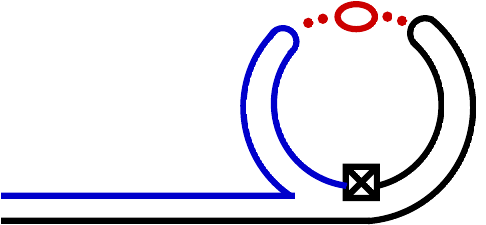}}
\caption{Quark flows for the NLO self-energy tadpoles (a-f) and
  current-vertex loops (g-i).  The $x$ and $y$ quarks are continuously
  connected to the external lines, closed loops are sea quarks, and
  current insertions are represented by crossed
  boxes.\label{fig:qflowdiag}}
\end{figure}
%
\section{\label{sec:mass}Next-to-leading order corrections to masses}
For a taste $t$ PGB $\phi^t_{xy}$ composed of quarks with flavors $x,y,\ x\neq y$,
the mass is defined in terms of the self-energy, as in continuum ChPT.  The NLO
mass can be obtained by adding the NLO self-energy to the tree-level mass,
\begin{align}
  M_{xy,\,t}^2 = m_{xy,\,t}^2 + \Sigma_{xy,\, t}(-m_{xy,\,t}^2) \,.
\end{align}
$\Sigma_{xy,\, t}$ consists of connected and disconnected tadpole loops with
vertices from the LO Lagrangian at $\mc{O}(\phi^4)$ and tree-level graphs with
vertices from the NLO Lagrangian at $\mc{O}(\phi^2)$.  The tadpole graphs
contribute the leading chiral logarithms, while the tree-level terms are
analytic in the quark masses and the square of the lattice spacing.  

We have not attempted to enumerate all terms in the NLO Lagrangian.  It
consists of generalizations of the Gasser-Leutwyler terms~\cite{Gasser:1984gg},
as in ordinary, unmixed SChPT, as well as generalizations of the Sharpe-Van de
Water Lagrangian~\cite{Sharpe:2004is} to the mixed action case.  There also
exist additional operators including traces over taste-singlets; such operators
vanish in the unmixed theory.  

Given the different kinds of operators in the NLO Lagrangian, the analytic
terms at NLO have the same form as those in the unmixed theory, but with
distinct LECs for valence-valence, sea-sea, and valence-sea PGBs.  
\jab{Explicitly, we have}
\begin{align}
\Sigma & ^\mr{NLO,\,anal}_{xy,\, t} (-m_{xy,\,t}^2) =
\frac{16}{f^2} \Big( ( 2 L_6 - L_4 ) \mu ( m_x + m_y ) \\
\times\ & 2 \mu ( m_u + m_d + m_s )
+ ( 2 L_8 - L_5 ) [ 2 \mu ( m_x + m_y ) ]^2 \Big) \nonumber \\
+\ & c_{1,t}^{xy} a^2 \mu ( m_x + m_y ) + c_{2,t}^{xy} a^2 2 \mu ( m_u + m_d + m_s
) \nonumber \\
+\ & c_{3,t}^{xy} a^4 \,, \nonumber
\end{align}
\jab{where the coefficients of the last three terms are linear combinations of
the LECs in the generalized Sharpe-Van de Water Lagrangian, and so depend on
the sector of the $x$ and $y$ quarks:  $c_{i,t}^{xy} = c_{i,t}^{vv},\
c_{i,t}^{v\sigma},\ c_{i,t}^{\sigma\sigma}$ for valence-valence, valence-sea,
and sea-sea mesons, respectively.  We note that for the NLO tree-level
diagrams, the symmetry between $x$ and $y$ quarks $x \leftrightarrow y$ is
present, as for the ordinary, unmixed case.}  

\jab{A few operators from the Sharpe-Van de Water Lagrangian suffice to justify
these claims.  We introduce projection operators onto the valence and sea
sectors in these operators, lift the degeneracy of the LECs in the three
sectors, and calculate the analytic contributions to the self-energies.  For
example, for first operator in Table~\ref{tab:nlo_lag}, we have}
\begin{align}
& a^2 C^{vv} \Tr ( P_v \partial_\mu \Sigma^{\dagger} \xi_5 P_v \partial_\mu \Sigma \xi_5 ) + ( v \rightarrow \sigma ) \\
+\ & a^2 C^{v\sigma} \left[ \Tr ( P_v \partial_\mu \Sigma^{\dagger} \xi_5 P_\sigma \partial_\mu \Sigma \xi_5 ) + p.c. \right] \,, \nonumber
\end{align}
\jab{which yield analytic contributions of the form}
\begin{align}
a^2 \theta^{t5} C^{xy} p^2 \,,
\end{align}
where the coefficient $C^{xy} = C^{vv},\ C^{\sigma\sigma},\ C^{v\sigma}$ for
$xy \in vv,\ \sigma\sigma,\ v\sigma$, and setting $p^2 = -\mu(m_x+m_y) -
a^2\Delta_t^{xy}$ yields terms like $c_{1,t}^{xy}$ and $c_{3,t}^{xy}$.

\begin{table}[htbp]
\begin{tabular}{lr}  \hline\hline
Operator & Order \\ \hline
$\Tr(\partial_\mu \Sigma^\dagger \xi_5 \partial_\mu \Sigma \xi_5)$ & $a^2 p^2$ \\
$\Tr(\xi_\mu \Sigma^\dagger \xi_\mu \Sigma^\dagger) \Tr(\Sigma M^\dagger) + p.c.$ & $a^2 m$ \\
$\Tr(\xi_5 \Sigma \xi_5 \Sigma^\dagger)\Tr(\xi_5 \Sigma \xi_5 \Sigma^\dagger)$ & $a^4$ \\
\hline\hline
\end{tabular}
\caption{
Examples of operators in the Sharpe-Van de Water Lagrangian contributing
tree-level terms to the masses and decay constants at NLO.  Such operators enter
with undetermined LECs that differ in the valence-valence, valence-sea, and
sea-sea sectors, breaking degeneracies between valence
and sea quarks.
\label{tab:nlo_lag}
}
\end{table}

Finally, we have calculated the tadpole graphs for the sea-sea PGBs and find
them identical to the results in the unmixed theory, as
expected~\cite{Bernard:1993sv}.  Below we consider the tadpole graphs for the
valence-valence and valence-sea PGBs.
\subsection{\label{subsec:val-val}Valence-valence sector}
For any $\Gamma_4\rtimes\,$SO(4) irrep, the calculation of the valence-valence
PGB self-energies proceeds as for the unmixed
case~\cite{Aubin:2003mg,SWME:2011aa}.  Quark flow diagrams corresponding to the
tadpole graphs are shown in diagrams (a-f) of Fig.~\ref{fig:qflowdiag}.  The
kinetic energy, mass, and $\mc{U}$ vertices yield graphs of types (a), (c), and
(d), and the taste-singlet potential vertices ($\propto C_\mr{mix}$) yield
graphs of type (a),
\begin{align}
\frac{a^2C_\mr{mix}}{3f^2(4\pi f)^2}\,\sum_{\val{i}^\prime\sea{i}^\prime b}\,\ell(m^2_{\val{i}^\prime \sea{i}^\prime,\,b})\,,\label{eq:Cmix}
\end{align}
where $\val{i}^\prime$ is summed over $\val{x},\val{y}$; $\sea{i}^\prime$ is
summed over the physical sea quarks $u,d,s$; and $\ell(m^2)\equiv m^2 \ln
(m^2/\Lambda^2) + \delta_1(mL)$ is the chiral logarithm, with $\Lambda$ the
scale of dimensional regularization and $\delta_1$ the correction for finite
spatial volume~\cite{Bernard:2001yj}. ($L$ is the spatial extent of the lattice.)

Vertices from $\mc{U}^\prime$ yield graphs of types (b), (e), and (f).  The
hairpin vertex graphs are of types (e) and (f).  As in the unmixed case, they
can be combined and eliminated in favor of a contribution of type (d).  In the
mixed-action case, the necessary identity is ($t\in V,A$)
\begin{align}
\frac{\delta^{vv}_F}{p^2+m_{\val{x}\val{x},\,t}^2} + \frac{\delta^{v\sigma}_F}{4}\sum_{\sea{i}^\prime} D^t_{\val{x}\sea{i}^\prime} = -(p^2+m_{\val{y}\val{y},\,t}^2) D^t_{\val{x}\val{y}}\,. \label{eq:coolID}
\end{align}
This relation follows from Eqs.~\eqref{eq:discDnotvv} and \eqref{eq:discDvv}.  

As in the unmixed theory, graphs of type (b) come from vertices $\propto
\omega^{vv}_t \equiv 16(C^{vv}_{2F} + C^{vv}_{5F})/f^2$ for $F=V,A$; they have
the same form as those in the unmixed case~\cite{SWME:2011aa}.

Adding the various contributions and evaluating the result at
$p^2=-m^2_{\val{x}\val{y},\,t}$, we have the NLO, one-loop contributions to the
self-energies of the valence-valence PGBs,
\begin{align}
-\Sigma&^{\text{NLO loop}}_{\val{x}\val{y},\,t}(-m^2_{\val{x}\val{y},\,t})=
\frac{a^2}{48(4\pi f)^2}\ \times \label{eq:Sigvv}\\
\sum_c\Biggl[&\left(\Delta^{vv,v\sigma}_{ct}-\Delta^{vv}_t-\Delta^{v\sigma}_c+\frac{16C_\mr{mix}}{f^2}\right)\,\sum_{\val{i}^\prime\sea{i}^\prime} \ell(m^2_{\val{i}^\prime\sea{i}^\prime,\,c})\nonumber\\
+\ &\frac{3}{2}\Biggl(\sum_{b\in V,A}\omega^{vv}_b\tau_{cbt}\tau_{cbt}(1+\theta^{ct})\Biggr)\ell(m^2_{\val{x}\val{y},\,c})\Biggr]\nonumber \\
+\ &\frac{1}{12(4\pi f)^2}\,\int\frac{d^4q}{\pi^2}\ \times \nonumber \\
\sum_c\Biggl[&a^2\left(\Delta^{vv}_{ct}-\Delta^{vv}_t-\Delta^{vv}_c\right)(D_{\val{xx}}^c+D_{\val{yy}}^c)\nonumber \\
+ \biggl[&\Bigl(2(1-\theta^{ct}) + \rho^{ct}\Bigr)q^2
+\Bigl(2(1+2\theta^{ct}) + \rho^{ct}\Bigr)m^2_{\val{x}\val{y},\,5}\nonumber \\
+\ &2a^2\Delta^{\prime vv}_{ct} + a^2\Bigl(2\theta^{ct}\Delta^{vv}_t + (2+\rho^{ct})\Delta^{vv}_c\Bigr)\biggr]D_{\val{x}\val{y}}^c\Biggr]\,,\nonumber
\end{align}
where $\rho^{ct}\equiv -4(2+\theta^{ct})$ unless $c=I$, when it vanishes, $\tau_{cbt}\equiv\Tr(T^cT^bT^t)$ is a trace over (a product of) generators of U(4), and 
\begin{align}
\Delta^{vv}_{ct}&\equiv\frac{8}{f^2}\sum_{b\neq I} C^{vv}_b(5+3\theta^{cb}\theta^{bt}-4\theta^{5b}\theta^{bt}-4\theta^{cb}\theta^{b5})\,,\\
\Delta^{\prime vv}_{ct}&\equiv\frac{8\theta^{ct}}{f^2}\sum_{b\neq I} C^{vv}_b(1+3\theta^{cb}\theta^{bt}-2\theta^{5b}\theta^{bt}-2\theta^{cb}\theta^{b5})\,,\\
\Delta^{vv,v\sigma}_{ct}&\equiv \frac{8}{f^2}\sum_{b\neq I}\bigl[\tfrac{1}{2}(9C^{vv}_b + C^{\sigma\sigma}_b) + C^{v\sigma}_b(3\theta^{cb}\theta^{bt}\nonumber\\
&-\ 4\theta^{cb}\theta^{b5})-4C^{vv}_b\theta^{5b}\theta^{bt}\bigr]\,.
\end{align}
The form of Eq.~\eqref{eq:Sigvv} is the same as that in ordinary
SChPT~\cite{SWME:2011aa}; the differences are in the definition of the
disconnected propagators and the LECs of the effective field theory.  The
reduction to the unmixed case is straightforward.

\jab{To illustrate the final results, we consider the pions of the 2+1 flavor theory
in the fully dynamical case.  The theory has two degenerate light quarks and
one strange quark in valence and sea sectors, with valence and sea quark masses
equal, for each flavor.  Substituting for the quark masses in
Eq.~\eqref{eq:Sigvv}, noting the degeneracies within each
$\Gamma_4\rtimes\,$SO(4) irrep, summing over the taste index $c$, and
performing the loop integrals, we have}
\begin{align}
-\Sigma&^{\text{NLO loop}}_{\pi^{vv}_t}(-m^2_{\pi^{vv}_t})=
\frac{a^2}{(4\pi f)^2}\ \times \label{eq:Sigvv_final} \\
\sum_B \biggl[ &\delta^{vv,vv}_{BF} \ell(\pi^{vv}_B) +\frac{\Delta^{vv,v\sigma}_{BF}}{24} \left( 2 \ell(\pi^{v\sigma}_B) + \ell(K^{v\sigma}_B) \right) \biggr] \nonumber \\
+\ &\frac{1}{12(4\pi f)^2}\,\times \nonumber \\
\Biggl[ & 2 \sum_X \left( 12 \pi_P - 6 \nu_{VF} X_V - a^2 (\Delta^{vv,vv}_{VF} + \Delta^{\prime\,vv,vv}_{VF}) \right) \nonumber\\
\times\ & (\delta^{v\sigma}_V)^2/\delta^{\sigma\sigma}_V \, \Bigl[ R^{S^{\sigma\sigma}}_{\pi^{vv}\eta^{\sigma\sigma}\eta^{\prime\,\sigma\sigma}}(X_V) \nonumber \\
+\ & a^2(\Delta^{\sigma\sigma}_V-\Delta^{vv}_V)D^{S^{\sigma\sigma}}_{\pi^{vv}\eta^{\sigma\sigma}\eta^{\prime\,\sigma\sigma},\pi^{vv}}(X_V) \Bigr] \ell(X_V) \nonumber \\
+\ & 2 \Bigl[(\delta^{v\sigma}_V)^2/\delta^{\sigma\sigma}_V \, a^2(\Delta^{\sigma\sigma}_V-\Delta^{vv}_V) R^{S^{\sigma\sigma}}_{\pi^{vv}\eta^{\sigma\sigma}\eta^{\prime\,\sigma\sigma}}(\pi^{vv}_V) \nonumber\\
+\ & \delta^{vv}_V - (\delta^{v\sigma}_V)^2/\delta^{\sigma\sigma}_V \Bigr]\Bigl( 6 \nu_{VF} \ell(\pi^{vv}_V) \nonumber \\
+\ &\left( 12 \pi_P - 6 \nu_{VF} \pi^{vv}_V - a^2 ( \Delta^{vv,vv}_{VF} + \Delta^{\prime\,vv,vv}_{VF} ) \right) \tilde{\ell}(\pi^{vv}_V) \Bigr) \nonumber \\
+\ & (V\rightarrow A) \nonumber \\
-\ & \tfrac{8}{3} \left( 3 \pi_P + 2 a^2 \Delta^{vv,vv}_{IF} \right) \biggl[ \sum_X \Bigl[ R^{S^{\sigma\sigma}}_{\pi^{vv}\eta^{\sigma\sigma}}(X_I) \nonumber \\
+\ & a^2( \Delta^{\sigma\sigma}_I - \Delta^{vv}_I ) D^{S^{\sigma\sigma}}_{\pi^{vv}\eta^{\sigma\sigma},\pi^{vv}}(X_I) \Bigr] \ell(X_I) \nonumber \\
+\ & a^2( \Delta^{\sigma\sigma}_I - \Delta^{vv}_I ) R^{S^{\sigma\sigma}}_{\pi^{vv}\eta^{\sigma\sigma}}(\pi^{vv}_I) \tilde{\ell}(\pi^{vv}_I) \biggr] \Biggr] \,. \nonumber
\end{align}
\jab{On the right side of Eq.~\eqref{eq:Sigvv_final}, we represent the squares of the tree-level masses by the names of the respective mesons,}
\begin{align}
\pi^{vv}_B &\equiv 2 \mu m_\ell + a^2 \Delta^{vv}_B \,, \\
\pi^{v\sigma}_B &\equiv 2 \mu m_\ell + a^2 \Delta^{v\sigma}_B \,, \\
K^{v\sigma}_B &\equiv \mu (m_\ell + m_s) + a^2 \Delta^{v\sigma}_B \,, \\
\pi_P &\equiv 2 \mu m_\ell \,, \\
S^{\sigma\sigma}_B &\equiv 2 \mu m_s + a^2 \Delta^{\sigma\sigma}_B \,,
\end{align}
\jab{and we define linear combinations of LECs that are degenerate within irreps of $\Gamma_4 \rtimes$SO(4).}
\begin{align}
\delta^{vv,vv}_{BF} &\equiv \sum_{c\in B} \frac{1}{32} \Biggl( \sum_{b\in V,A} \omega^{vv}_b \tau_{cbt} \tau_{cbt} ( 1 + \theta^{ct} ) \Biggr ) \,, \\
\Delta^{vv,v\sigma}_{BF} &\equiv \sum_{c\in B} \left( \Delta^{vv,v\sigma}_{ct} - \Delta^{vv}_t - \Delta^{v\sigma}_c + \frac{16C_\mr{mix}}{f^2} \right) \,, \\
\Delta^{vv,vv}_{BF} &\equiv \sum_{c\in B} \left( \Delta^{vv}_{ct} - \Delta^{vv}_t - \Delta^{vv}_c \right) \,, \\
\Delta^{\prime vv,vv}_{BF} &\equiv \sum_{c\in B} \left( \Delta^{\prime vv}_{ct} + \Bigl( \theta^{ct} \Delta^{vv}_t + ( 1 + \rho^{ct}/2 ) \Delta^{vv}_c \Bigr) \right) \,.
\end{align}
\jab{The whole number $\nu_{BF} \equiv \ohf \sum_{c\in B} ( 1 + \theta^{ct} )$
counts the taste matrices in irrep $B$ commuting with the taste matrix $\xi_t$,
where $t\in F$.}
\jab{The index of summation $X$ runs over the meson names in the subscripts of the
residues, which are defined as in Ref.~\cite{SWME:2011aa},}
\begin{align}
R^{A_1 A_2 \cdots A_k}_{B_1 B_2 \cdots B_n}(X_F) &\equiv \frac{ \prod_{A_{jF}} ( A_{jF} - X_F )
}{ \prod_{ B_{iF} \neq X_F } ( B_{iF} - X_F ) }\,, \\
D^{A_1 A_2 \cdots A_k}_{B_1 B_2 \cdots B_n, B_i}(X_F) &\equiv -
\frac{\partial}{\partial B_{iF}} R^{A_1 A_2 \cdots A_k}_{B_1 B_2 \cdots
B_n}(X_F) \,.
\end{align}
\jab{The chiral behavior of the mixed action theory differs nontrivially from that
of the unmixed theory due to incomplete cancellation of double poles in the
loop integrals.  The chiral logarithm $\tilde{\ell}(m^2)\equiv
-(\ln(m^2/\Lambda^2) + 1) + \delta_3(mL)$, with the finite volume correction
$\delta_3$~\cite{Bernard:2001yj}, arises from these loops.  Unlike in the
ordinary theory, such terms enter here even though valence and sea quark masses
for each flavor are equal, \textit{i.e.}, in the fully dynamical case.}

The valence-valence, taste-pseudoscalar PGBs are true Goldstone bosons in the
chiral limit, $m_x,m_y\rightarrow 0,\ a\neq 0$.  Setting $t=5$ in
Eq.~\eqref{eq:Sigvv} and noting that
\begin{align}
\Delta^{vv,v\sigma}_{c5}&=\Delta^{v\sigma}_c - \frac{16C_\mr{mix}}{f^2}\label{eq:Delmix-vsig}\\
\Delta^{vv}_{c5}&=\Delta^{vv}_c\\
\Delta^{\prime vv}_{c5}&=-\theta^{c5}\Delta^{vv}_c\\
\Delta^{vv}_5&=0\,,
\end{align}
we have
\begin{align}
-\Sigma^{\text{NLO loop}}_{\val{x}\val{y},\,5}(-m^2_{\val{x}\val{y},\,5})=\frac{\mu(m_{\val{x}}+m_{\val{y}})}{2(4\pi f)^2}\,\sum_b\theta^{b5}\int\frac{d^4q}{\pi^2}D^b_{\val{x}\val{y}}\,,\label{eq:gold}
\end{align}
which is the generalization of the results of Ref.~\cite{Aubin:2003mg} to the
mixed-action case.  As in ordinary SChPT, only graphs of type (d) contribute.
To generalize to the mixed-action theory, one has only to replace the
disconnected propagators $D^t_{\val{x}\val{y}}$ with their counterparts in the
mixed-action theory.  

%

\subsection{\label{subsec:val-sea}Valence-sea sector}
We consider mesons $\phi^t_{\val{x}\sea{y}}$ with one valence quark $\val{x}$ and one sea quark $\sea{y}$.   For tadpoles with vertices from the kinetic energy and mass terms of the LO Lagrangian (Eq.~\eqref{eq:LOlag}), we find graphs of types (a), (c), and (d).
\begin{align}
\frac{1}{48(4\pi f)^2}\,\sum_{c,\sea{i}^\prime}\Bigg[
  \left(
  p^2 + \mu(m_{\val{x}} + m_{\sea{y}}) - a^2\Delta^{v\sigma}_c
  \right)
  \ell(m^2_{\val{x}\sea{i}^\prime,c})  \\
  +\ \left(
  p^2 + \mu(m_{\val{x}} + m_{\sea{y}}) - a^2\Delta^{\sigma\sigma}_c
  \right)
  \ell(m^2_{\sea{y}\sea{i}^\prime,c}) \Bigg] \nonumber
\end{align}
\begin{align}
+\ \frac{1}{12(4\pi f)^2}\,\sum_c \int \frac{d^4q}{\pi^2}
  \Biggl[
  \left(
  p^2 + q^2 + \mu(3m_{\val{x}} + m_{\sea{y}})
  \right) D^c_{\val{x}\val{x}}
  \nonumber\\
  +\ \left(
  p^2 + q^2 + \mu(m_{\val{x}} + 3m_{\sea{y}})
  \right) D^c_{\sea{y}\sea{y}}
  \nonumber\\
  -\ 2\theta^{ct} \left(
  p^2 + q^2 - \mu(m_{\val{x}} + m_{\sea{y}})
  \right) D_{\val{x}\sea{y}}^c
  \Biggr]\,,\nonumber
\end{align}
where $\sea{i}^\prime$ is summed over the physical sea-quark flavors.
As for the sea-sea and valence-valence sectors, the $q^2 D^c_{\val{x}\val{x}}$ and $q^2 D^c_{\sea{y}\sea{y}}$ terms can be eliminated in favor of a $q^2 D^c_{\val{x}\sea{y}}$ term.  But for the valence-sea mesons, an additional term arises, with the form of a connected contribution [graph (e) of Fig.~\ref{fig:qflowdiag}].  The necessary identities are
\begin{align}
(q^2 + 2\mu m_{\val{x}})D^t_{\val{x}\val{x}} &= \frac{\delta^{v\sigma}_F}{\delta^{\sigma\sigma}_F}(q^2 + m_{\sea{yy},\,t}^2) D^t_{\val{x}\sea{y}} \label{eq:vsig_q2Dxx_id}\\
 &-\, a^2\Delta^{vv}_F D^t_{\val{x}\val{x}}
 + \frac{(\delta^{v\sigma}_F)^2/\delta^{\sigma\sigma}_F - \delta^{vv}_F}{q^2 + m_{\val{xx},\,t}^2}\,,\nonumber \\
(q^2 + 2\mu m_{\sea{y}})D^t_{\sea{y}\sea{y}} &= \frac{\delta^{\sigma\sigma}_F}{\delta^{v\sigma}_F}(q^2 + m_{\val{xx},\,t}^2) D^t_{\val{x}\sea{y}} \label{eq:vsig_q2Dyy_id} \\
 &-\, a^2\Delta^{\sigma\sigma}_F D^t_{\sea{y}\sea{y}}\,,\nonumber
\end{align}
which hold for $t\in F = V,A,I$.  Applying these identities to the above result
gives
\begin{align}
\frac{1}{48(4\pi f)^2}\,\sum_{c,\sea{i}^\prime}\Bigg[&
  \left(
  p^2 + \mu(m_{\val{x}} + m_{\sea{y}}) - a^2\Delta^{v\sigma}_c
  \right)
  \ell(m^2_{\val{x}\sea{i}^\prime,c}) \\ 
  +\ &\left(
  p^2 + \mu(m_{\val{x}} + m_{\sea{y}}) - a^2\Delta^{\sigma\sigma}_c
  \right)
  \ell(m^2_{\sea{y}\sea{i}^\prime,c}) \Bigg] \nonumber \\
  -\ \frac{1}{12(4\pi f)^2}&\sum_{c\in V,A} 
  \left(
  \delta^{vv}_c - (\delta^{v\sigma}_c)^2/\delta^{\sigma\sigma}_c \right)
  \ell(m^2_{\val{x}\val{x},c}) \nonumber \\
+\ \frac{1}{12(4\pi f)^2}&\sum_c \int \frac{d^4q}{\pi^2}\ \times \nonumber \\
  \Biggl[
  &\left(
  p^2 + \mu(m_{\val{x}} + m_{\sea{y}}) - a^2\Delta^{vv}_c
  \right) D^c_{\val{x}\val{x}}
  \nonumber\\
  +\ &\left(
  p^2 + \mu(m_{\val{x}} + m_{\sea{y}}) - a^2\Delta^{\sigma\sigma}_c
  \right) D^c_{\sea{y}\sea{y}}
  \nonumber \\
  +\ &\Biggl(
  -2\theta^{ct} p^2 + \left( \frac{\delta^{v\sigma}_c}{\delta^{\sigma\sigma}_c} + \frac{\delta^{\sigma\sigma}_c}{\delta^{v\sigma}_c} - 2\theta^{ct} \right) q^2 \nonumber \\
  + \left( \frac{\delta^{\sigma\sigma}_c}{\delta^{v\sigma}_c} + \theta^{ct} \right) & (2\mu m_{\val{x}})
  + \left( \frac{\delta^{v\sigma}_c}{\delta^{\sigma\sigma}_c} + \theta^{ct} \right)(2\mu m_{\sea{y}}) \nonumber \\
  +\ &a^2\left( \frac{\delta^{v\sigma}_c}{\delta^{\sigma\sigma}_c} \Delta^{\sigma\sigma}_c + \frac{\delta^{\sigma\sigma}_c}{\delta^{v\sigma}_c} \Delta^{vv}_c \right)
  \Biggr) D_{\val{x}\sea{y}}^c
  \Biggr]\,.\nonumber
\end{align}

From the taste-singlet potential, we find contributions not only from graphs of type (a), as in the valence-valence sector, but also from graphs of types (c) and (d),
\begin{align}
\frac{a^2C_\mr{mix}}{3f^2(4\pi f)^2}\,\sum_{b}\Bigg[\sum_{\sea{i}^\prime}
  \left(
  8\ell(m^2_{\val{x}\sea{i}^\prime,b}) + \ell(m^2_{\sea{y}\sea{i}^\prime,b}) 
  \right)
  \label{eq:vsig_Ipot}\\
+\ 4\int \frac{d^4q}{\pi^2}
  \left(
  D^b_{\val{x}\val{x}} + D^b_{\sea{y}\sea{y}} - 2\theta^{bt} D_{\val{x}\sea{y}}^b
  \right)
  \Bigg]\,.\nonumber
\end{align}

From the single-trace potential $\mc{U}$, we have graphs of types (a), (c), and (d),
\begin{align}
  \frac{a^2}{48(4\pi f)^2}
  \sum_{b}\Bigg[\sum_{\sea{i}^\prime}\Big(
  \Delta_{bt}^{v\sigma,v\sigma}
  \ell(m^2_{\val{x}\sea{i}^\prime,b}) \label{eq:vsig_U} \\
  +\
  \Delta_{bt}^{v\sigma,\sigma\sigma}
  \ell(m^2_{\sea{y}\sea{i}^\prime,b})
  \Big) \nonumber \\
  +\ 4 \int \frac{d^4q}{\pi^2}
  \Big(
  \Delta_{bt}^{v\sigma,vv}
  D^b_{\val{x}\val{x}}
  +
  \Delta_{bt}^{v\sigma,\sigma\sigma}
  D^b_{\sea{y}\sea{y}} \nonumber \\
  +\
  2 \Delta_{bt}^{\prime\, v\sigma,v\sigma}
  D_{\val{x}\sea{y}}^b
  \Big)
  \Bigg]\,,\nonumber
\end{align}
where
\begin{align}
  \Delta_{ct}^{v\sigma,v\sigma}
  \equiv \frac{8}{f^2}\sum_{b\neq I} &
  \bigl[
  4C^{vv}_b
  + C^{\sigma\sigma}_b \bigl( 1
  + 3 \theta^{cb} \theta^{bt} \bigr)
  \\
  -\ &4 C^{v\sigma}_b \bigl( \theta^{5b}\theta^{bt}
  + \theta^{cb}\theta^{b5} \bigr)
  \bigr]\,,\nonumber
  \\
  \Delta_{ct}^{v\sigma,\sigma\sigma}
  \equiv \frac{8}{f^2}\sum_{b\neq I} &
  \bigl[
  C^{\sigma\sigma}_b \bigl( \tfrac{9}{2} 
  - 4 \theta^{cb}\theta^{b5} \bigr)
  + \tfrac{1}{2}C^{vv}_b  \\
  +\ & C^{v\sigma}_b \bigl( 3 \theta^{cb} \theta^{bt}
  - 4 \theta^{5b}\theta^{bt} \bigr)
  \bigr]\,, \nonumber \\
  \Delta_{ct}^{v\sigma,vv}
  \equiv \frac{8}{f^2}\sum_{b\neq I} &
  \bigl[
  C^{vv}_b \bigl( \tfrac{9}{2} 
  - 4 \theta^{cb}\theta^{b5} \bigr)
  + \tfrac{1}{2}C^{\sigma\sigma}_b  \\
  +\ & C^{v\sigma}_b \bigl( 3 \theta^{cb} \theta^{bt} - 4 \theta^{5b}\theta^{bt} \bigr) \bigr]\,,\nonumber\\
  \Delta_{ct}^{\prime\,v\sigma,v\sigma}
  \equiv \frac{8\theta^{ct}}{f^2}\sum_{b\neq I} &
  \bigl[
    \bigl( C^{vv}_b
  + C^{\sigma\sigma}_b \bigr) \bigl( \tfrac{1}{2}
  - \theta^{cb}\theta^{b5} \bigr)  \\
  +\ & C^{v\sigma}_b \bigl( 3 \theta^{cb} \theta^{bt}
  - 2 \theta^{5b}\theta^{bt} \bigr)
  \bigr]\,.\nonumber
\end{align}
In the unmixed case,
$\Delta_{ct}^{v\sigma,v\sigma}=\Delta_{ct}^{v\sigma,\sigma\sigma}=\Delta_{ct}^{v\sigma,vv}=\Delta_{ct}$,
$\Delta_{ct}^{\prime\,v\sigma,v\sigma}=\Delta_{ct}^{\prime}$, and the
contribution from $\mc{U}$ reduces \cite{Aubin:2003mg,SWME:2011aa}.  We note
that $\Delta_{ct}^{v\sigma,\sigma\sigma}$ appears in both connected and
disconnected terms.

From the double-trace potential $\mc{U}^\prime$, we have, after combining graphs of types (e) and (f) to eliminate those of type (f),
\begin{align}
&\frac{1}{12(4\pi f)^2}\ \times \\
\sum_c \biggl[&\frac{3a^2}{8} \sum_{b\in V,A}
    \tau_{cbt}\tau_{cbt} 
    \left( \omega^{v\sigma}_b + \frac{\theta^{ct}}{2}
                                ( \omega^{vv}_b + \omega^{\sigma\sigma}_b ) 
    \right)
    \ell ( m_{\val{x}\sea{y},\,c}^2 ) \nonumber
\\
+ &\int \frac{d^4q}{\pi^2} \rho^{ct}
    \left( q^2 + m_{\val{x}\sea{y},\,5}^2 + \frac{a^2}{2}
                                ( \Delta^{vv}_c + \Delta^{\sigma\sigma}_c )
    \right)
    D^c_{\val{x}\sea{y}} \biggr] \nonumber
\\
+\ &\frac{1}{3(4\pi f)^2}\sum_{c\in V,A} \biggl[ \Bigl( \delta^{vv}_c 
- ( \delta^{v\sigma}_c )^2 / \delta^{\sigma\sigma}_c \Bigr) \ell(m^2_{\val{xx},c}) \nonumber
\\
+ &\int \frac{d^4q}{\pi^2} \biggl( \Bigl( 2 - \delta^{\sigma\sigma}_c / \delta^{v\sigma}_c
                                       - \delta^{v\sigma}_c / \delta^{\sigma\sigma}_c \Bigr) q^2 \nonumber
\\
+\ &\Bigl( 1 - \delta^{\sigma\sigma}_c / \delta^{v\sigma}_c \Bigr) ( 2 \mu m_{\val{x}} + a^2 \Delta^{vv}_c ) \nonumber
\\
+\ &\Bigl( 1 - \delta^{v\sigma}_c / \delta^{\sigma\sigma}_c \Bigr) ( 2 \mu m_{\sea{y}} + a^2 \Delta^{\sigma\sigma}_c ) \biggr)
D^c_{\val{x}\sea{y}}
\biggr]\,.\nonumber
\end{align}
The reduction of this expression in the unmixed case is immediate.  In the
valence-valence sector and unmixed cases, the graphs of types (e) and (f) can
be combined into a graph of type (d).  In the valence-sea sector, we eliminate
graphs of type (f) in favor of those of type (d), but a contribution of type
(e) remains.

Adding the various contributions and evaluating the sum at
$p^2=-m^2_{\val{x}\sea{y},\,t}$ gives, for graphs with connected propagators, 
\begin{align}
  -\Sigma&_{\val{x}\sea{y},\,t}^{\mr{NLO\ loop,\ con}}(-m^2_{\val{x}\sea{y},t})
  = \frac{a^2}{48(4\pi f)^2}\ \times \label{eq:Sigvs-con}\\
 \sum_c
  \Bigg[
  & \left(
  \Delta_{ct}^{v\sigma,v\sigma} - \Delta^{v\sigma}_t -\Delta^{v\sigma}_c  +
  \frac{128 C_\text{mix}}{f^2}\right)
  \sum_{\sea{i}^\prime} \ell(m^2_{\val{x}\sea{i}^\prime,c})  \nonumber \\
  +\ &\left(
  \Delta_{ct}^{v\sigma,\sigma\sigma} - \Delta^{v\sigma}_t -\Delta^{\sigma\sigma}_c  +
  \frac{16 C_\text{mix}}{f^2}\right)
  \sum_{\sea{i}^\prime} \ell(m^2_{\sea{y}\sea{i}^\prime,c}) \nonumber \\
  +\ &\frac{3}{2}\sum_{b\in V,A}
  \tau_{cbt}\tau_{cbt}
  \left(\omega_b^{v\sigma}+
  \frac{\theta^{ct}}{2}(\omega_b^{vv}+\omega_b^{\sigma\sigma})
  \right)\ell(m^2_{\val{x}\sea{y},c})
  \Bigg] \nonumber \\
  +\ &\frac{1}{4(4\pi f)^2}\sum_{c\in V,A} \left(\delta^{vv}_c - (\delta^{v\sigma}_c)^2/\delta^{\sigma\sigma}_c\right)\ell(m^2_{\val{x}\val{x},c})\,, \nonumber
\end{align}
while for the graphs with disconnected propagators, we have
\begin{align}
  -\Sigma&_{\val{x}\sea{y},\,t}^{\mr{NLO\ loop,\ disc}}(-m^2_{\val{x}\sea{y},t})
  =\ \frac{1}{12(4\pi f)^2}\int\frac{d^4q}{\pi^2}\ \times \label{eq:Sigvs-disc} \\
  \sum_{c}
  \Biggl[
  & a^2\left(
  \Delta_{ct}^{v\sigma,vv}-\Delta^{v\sigma}_t-\Delta^{vv}_c +
  \frac{16 C_\text{mix}}{f^2}\right) D^c_{\val{x}\val{x}}
  \nonumber\\
  +\ & a^2\left(
  \Delta_{ct}^{v\sigma,\sigma\sigma}-\Delta^{v\sigma}_t-\Delta^{\sigma\sigma}_c +
  \frac{16 C_\text{mix}}{f^2}\right)
  D^c_{\sea{y}\sea{y}}
  \nonumber\\
  +\ & \bigg[
  \left(8-3\bigg(\frac{\delta^{v\sigma}_c}{\delta^{\sigma\sigma}_c}+\frac{\delta^{\sigma\sigma}_c}{\delta^{v\sigma}_c}\bigg)-2\theta^{ct}+\rho^{ct}\right)q^2 \nonumber \\
  +\ & \left( 4 - 3 \frac{ \delta^{\sigma\sigma}_c }{ \delta^{v\sigma}_c } + 2 \theta^{ct} + \frac{ \rho^{ct} }{2} \right)( 2 \mu m_{\val{x}} ) \nonumber \\
  +\ & \left( 4 - 3 \frac{ \delta^{v\sigma}_c }{ \delta^{\sigma\sigma}_c } + 2 \theta^{ct} + \frac{ \rho^{ct} }{2} \right)( 2 \mu m_{\sea{y}} ) + 2 a^2\Delta_{ct}^{\prime\, v\sigma,v\sigma} \nonumber \\
  +\ & a^2\bigg(2\theta^{ct}\Delta^{v\sigma}_t + \bigg(4-3\frac{\delta^{\sigma\sigma}_c}{\delta^{v\sigma}_c}+\frac{\rho^{ct}}{2}\bigg)\Delta^{vv}_c \nonumber \\
  +\ & \bigg(4-3\frac{\delta^{v\sigma}_c}{\delta^{\sigma\sigma}_c}+\frac{\rho^{ct}}{2}\bigg)\Delta^{\sigma\sigma}_c \bigg)
  - \frac{32a^2\theta^{ct}C_\text{mix}}{f^2}
  \bigg]D_{\val{x}\sea{y}}^c
  \Biggr]\,.\nonumber
\end{align}
The reduction in the unmixed case is straightforward.  There is no symmetry
under $\val{x}\leftrightarrow \sea{y}$; when using the replica method, the
valence and sea sectors of the effective theory are distinguished by the
operations of partial quenching (the valence quarks) and rooting (the sea
quarks).  
The taste-pseudoscalars are not Goldstone bosons (in the chiral limit) at
non-zero lattice spacing, and the self-energy does not vanish in the chiral
limit.
In the continuum limit, the symmetry is restored, and the masses vanish, in
accord with Goldstone's theorem.

\jab{To illustrate the final results, we again consider the pions of the 2+1
flavor theory with degenerate valence and sea quarks.  We have}
\begin{align}
  -\Sigma&_{\pi^{v\sigma}_t}^{\mr{NLO\ loop}}(-m^2_{\pi^{v\sigma}_t})
  = \frac{a^2}{(4\pi f)^2}\ \times \label{eq:Sigvs_final} \\
 \sum_B
  \Bigg[
  & \delta^{v\sigma,v\sigma}_{BF} \ell(\pi^{v\sigma}_B) + \frac{\Delta^{v\sigma,v\sigma}_{BF}}{48} \left( 2 \ell(\pi^{v\sigma}_B) + \ell(K^{v\sigma}_B) \right) \nonumber \\
+\ &\frac{\Delta^{v\sigma,\sigma\sigma}_{BF}}{48}\left( 2 \ell(\pi^{\sigma\sigma}_B) + \ell(K^{\sigma\sigma}_B) \right) 
  \Bigg] \nonumber \\
+\ &\frac{1}{(4\pi f)^2} \left(\delta^{vv}_V - (\delta^{v\sigma}_V)^2/\delta^{\sigma\sigma}_V\right)\ell(\pi^{vv}_V) + (V\rightarrow A) \nonumber \\
+\ &\frac{1}{12(4\pi f)^2}\ \times \nonumber \\
\Bigg\{ \delta^{v\sigma}_V \sum_X &\Biggl[ 2 \Biggl( 6 \left( \frac{\delta^{\sigma\sigma}_V}{\delta^{v\sigma}_V} + \frac{\delta^{v\sigma}_V}{\delta^{\sigma\sigma}_V} \right) \pi_P - 6\biggl[ \nu_{VF} \nonumber \\
+\ &\left( \frac{\delta^{\sigma\sigma}_V}{\delta^{v\sigma}_V} + \frac{\delta^{v\sigma}_V}{\delta^{\sigma\sigma}_V} - 2 \right) \biggr] X_V - a^2 \bigg( \frac{\Delta^{v\sigma,vv}_{VF}}{2} \frac{\delta^{v\sigma}_V}{\delta^{\sigma\sigma}_V} \nonumber \\
+\ &\Delta^{\prime v\sigma,v\sigma}_{VF} \bigg)
\Biggr) 
R^{S^{\sigma\sigma}}_{\pi^{vv}\eta^{\sigma\sigma}\eta^{\prime\,\sigma\sigma}}(X_V)
- \frac{\delta^{v\sigma}_V}{\delta^{\sigma\sigma}_V} a^2 \Delta^{v\sigma,vv}_{VF} \nonumber \\
\times\ &a^2 \left(\Delta^{\sigma\sigma}_V - \Delta^{vv}_V \right)
D^{S^{\sigma\sigma}}_{\pi^{vv}\eta^{\sigma\sigma}\eta^{\prime\,\sigma\sigma},\pi^{vv}}(X_V)
\Biggr] \ell(X_V) \nonumber \\
-\ &a^2 \Delta^{v\sigma,\sigma\sigma}_{VF} \delta^{\sigma\sigma}_V
\sum_X R^{S^{\sigma\sigma}}_{\pi^{\sigma\sigma}\eta^{\sigma\sigma}\eta^{\prime\,\sigma\sigma}} (X_V) \ell(X_V) \nonumber \\
-\ & a^2 \Delta^{v\sigma,vv}_{VF} \Big[ (\delta^{v\sigma}_V)^2/\delta^{\sigma\sigma}_V a^2 (\Delta^{\sigma\sigma}_V - \Delta^{vv}_V ) \nonumber \\
\times\ &R^{S^{\sigma\sigma}}_{\pi^{vv}\eta^{\sigma\sigma}\eta^{\prime\,\sigma\sigma}}(\pi^{vv}_V)
+ \delta^{vv}_V - (\delta^{v\sigma}_V)^2/\delta^{\sigma\sigma}_V \Big] \tilde{\ell}(\pi^{vv}_V) \nonumber \\
+\ &(V\rightarrow A) \nonumber \\
-\ &\frac{4}{3}\Bigg[ \left( 2 ( 3 \pi_P + a^2 \Delta^{v\sigma,vv}_{IF} ) + a^2 \Delta^{v\sigma,\sigma\sigma}_{IF} \right) \nonumber \\
\times\ &\sum_X R^{S^{\sigma\sigma}}_{\pi^{vv}\eta^{\sigma\sigma}}(X_I)\ell(X_I) \nonumber \\
+\ &a^2 \Delta^{v\sigma,vv}_{IF} a^2(\Delta^{\sigma\sigma}_I - \Delta^{vv}_I) \Big[ R^{S^{\sigma\sigma}}_{\pi^{vv}\eta^{\sigma\sigma}}(\pi^{vv}_I) \tilde{\ell}(\pi^{vv}_I) \nonumber \\
+\ &\sum_X D^{S^{\sigma\sigma}}_{\pi^{vv}\eta^{\sigma\sigma},\pi^{vv}}(X_I)\ell(X_I)\Big] \nonumber \\
+\ & a^2\Delta^{v\sigma,\sigma\sigma}_{IF} \sum_X R^{S^{\sigma\sigma}}_{\pi^{\sigma\sigma}\eta^{\sigma\sigma}}(X_I)\ell(X_I) \nonumber 
\Bigg]
\Bigg\}\,.
\end{align}
\jab{The new linear combinations of LECs are}
\begin{align}
\delta^{v\sigma,v\sigma}_{BF} &\equiv \frac{1}{32} \sum_{c\in B} \sum_{b \in V,A} \tau^2_{cbt} \left( \omega^{v\sigma}_b + \ohf \theta^{ct} ( \omega^{vv}_b + \omega^{\sigma\sigma}_b ) \right) \,, \\
\Delta^{v\sigma,v\sigma}_{BF} &\equiv \sum_{c\in B} \left( \Delta^{v\sigma,v\sigma}_{ct} - \Delta^{v\sigma}_t - \Delta^{v\sigma}_c + \frac{ 128 C_\mr{mix} }{ f^2 } \right) \,, \\
\Delta^{v\sigma,\sigma\sigma}_{BF} &\equiv \sum_{c\in B} \left( \Delta^{v\sigma,\sigma\sigma}_{ct} - \Delta^{v\sigma}_t - \Delta^{\sigma\sigma}_c + \frac{ 16 C_\mr{mix} }{ f^2 } \right) \,, \\
\Delta^{v\sigma,vv}_{BF} &\equiv \sum_{c\in B} \left( \Delta^{v\sigma,vv}_{ct} - \Delta^{v\sigma}_t - \Delta^{vv}_c + \frac{ 16 C_\mr{mix} }{ f^2 } \right) \,, \\
\Delta^{\prime v\sigma,v\sigma}_{BF} &\equiv \sum_{c\in B} \bigg[ \Delta^{\prime v\sigma,v\sigma}_{ct} + \theta^{ct} \Delta^{v\sigma}_t + \frac{\Delta^{vv}_c}{2} \bigg( 4 - 3 \frac{\delta^{\sigma\sigma}_c}{\delta^{v\sigma}_c} \nonumber \\
+\ &\frac{\rho^{ct}}{2} \bigg) + \frac{\Delta^{\sigma\sigma}_c}{2} \bigg( 4 - 3 \frac{\delta^{v\sigma}_c}{\delta^{\sigma\sigma}_c} + \frac{\rho^{ct}}{2} \bigg) - \frac{ 16 \theta^{ct} C_\mr{mix} }{f^2} \bigg] \,,
\end{align} 
\jab{and we use the identity $\Delta^{\prime v\sigma,v\sigma}_{IF} = \ohf (
\Delta^{v\sigma,vv}_{IF} + \Delta^{v\sigma,\sigma\sigma}_{IF} )$ to simplify
the disconnected loops in the taste singlet channel.  As for the
valence-valence masses, the chiral behavior differs nontrivially from that of
the ordinary unmixed theory.  Even in the fully dynamical theory, double poles
do not completely cancel from the loop integrals.}
%
%
\section{\label{sec:decay}Next-to-leading order corrections to decay constants}
As for continuum and ordinary SChPT, the decay constants are defined by matrix
elements of the axial currents,
\begin{align}
  -i f_{xy,\,t}\, p_\mu = \bra{0}\, j^{\mu5}_{xy,\,t}\, \ket{\phi^t_{xy}(p)} \,.
\end{align}
The NLO corrections are the same types of diagrams that appear in continuum and
unmixed SChPT.  We have one-loop wave function renormalization contributions
[graphs (a), (c), and (d) of Fig.~\ref{fig:qflowdiag}], one-loop graphs from
insertions of the $\mc{O}(\phi^3)$-terms of the LO current [graphs (g), (h),
and (i) of Fig.~\ref{fig:qflowdiag}], and terms analytic in the quark masses
and squared lattice spacing, from the NLO Lagrangian~\cite{Aubin:2003uc}.  As
for the NLO analytic corrections to the masses, the NLO analytic corrections to
the decay constants have the same form as in the unmixed theory, with distinct
LECs for the valence-valence, sea-sea, and valence-sea sectors.

Turning to the one-loop corrections, we note that the LO current is determined
by the kinetic energy vertices of the LO Lagrangian; these vertices are the
same in mixed-action and unmixed SChPT.  Therefore, the LO current in the
mixed-action case is the same as the LO current in unmixed SChPT.  Likewise,
the NLO wave function renormalization corrections are determined by self-energy
contributions from tadpoles with kinetic energy vertices from the LO
Lagrangian.  Moreover, nothing in the calculation of the relevant part of the
self-energies or the current-vertex loops is sensitive to the sector of the
external quarks.  

Therefore, to generalize the one-loop graphs of the unmixed case, we have only
to replace the propagators with those of the mixed-action theory.  The results
hold for all sectors of the mixed-action theory (valence-valence, sea-sea, and
valence-sea).  \jab{Including the analytic contributions, w}e have
\begin{align}
&\frac{f^\text{NLO}_{xy,\,t}}{f} = 1 - \frac{1}{8(4\pi f)^2}\sum_c\ \times \label{eq:decayfin} \\
&\left[\frac{1}{4}\sum_{\val{i}^\prime\sea{i}^\prime} \ell(m^2_{\val{i}^\prime\sea{i}^\prime,\,c}) + \int\frac{d^4q}{\pi^2}(D^c_{xx}+D^c_{yy}-2\theta^{ct}D^c_{xy})\right] \nonumber \\
&+ \frac{16}{f^2} L_4 \mu ( m_u + m_d + m_s )
+ \frac{8}{f^2} L_5 \mu ( m_x + m_y ) + a^2 c^{xy}_t \,, \nonumber
\end{align}
\jab{where the coefficient $c^{xy}_t = c^{vv}_t,\ c^{v\sigma}_t,\
c^{\sigma\sigma}_t$ for valence-valence, valence-sea, and sea-sea mesons,
respectively.}
The form of this result is the same as that in the unmixed
theory~\cite{Bailey:2012jy}, and the reduction in the unmixed case is
immediate.  
\jab{As for the masses, the form of the NLO analytic terms can be verified by
considering a few operators in the generalized Sharpe-Van de Water Lagrangian
and calculating the resulting contributions.  In addition to the wave function
renormalization contributions, there are those from the NLO current.  But the
latter cannot change the form of the results, and considering the wave function
renormalization suffices.}

\jab{To illustrate the loop corrections, we begin with the valence-valence pions
in the 2+1 flavor, fully dynamical theory.  We have}
\begin{align}
&\frac{f^\text{NLO loop}_{\pi^{vv}_t}}{f} = - \frac{1}{16(4\pi f)^2} 
\sum_B g_B \left( 2 \ell(\pi^{v\sigma}_B) + \ell(K^{v\sigma}_B) \right) 
\label{eq:decay-vv-pion_fin} \\
+\ &\frac{(1-\Theta^{VF}/4)}{(4\pi f)^2} \Bigg[ \sum_X (\delta^{v\sigma}_V)^2/\delta^{\sigma\sigma}_V \bigg(  R^{S^{\sigma\sigma}}_{\pi^{vv}\eta^{\sigma\sigma}\eta^{\prime\,\sigma\sigma}} (X_V) \nonumber \\
+\ &a^2 (\Delta^{\sigma\sigma}_V - \Delta^{vv}_V) D^{S^{\sigma\sigma}}_{\pi^{vv}\eta^{\sigma\sigma}\eta^{\prime\,\sigma\sigma},\pi^{vv}} (X_V) 
\bigg) \ell(X_V) \nonumber \\
+\ &\Big[ (\delta^{v\sigma}_V)^2/\delta^{\sigma\sigma}_V a^2 ( \Delta^{\sigma\sigma}_V - \Delta^{vv}_V ) R^{S^{\sigma\sigma}}_{\pi^{vv}\eta^{\sigma\sigma}\eta^{\prime\,\sigma\sigma}}(\pi^{vv}_V) \nonumber \\
+\ &\left( \delta^{vv}_V - ( \delta^{v\sigma}_V )^2/ \delta^{\sigma\sigma}_V \right)
\Big] \tilde{\ell}(\pi^{vv}_V)
\Bigg] + (V\rightarrow A)\,, \nonumber
\end{align}
\jab{where $g_B \equiv \sum_{c \in B} 1$ and $\Theta^{BF} \equiv \sum_{c \in B}
\theta^{ct}$, as for the unmixed case.}
\jab{For the valence-sea pions in the 2+1 flavor, fully dynamical theory, we
have}
\begin{align}
&\frac{f^\text{NLO loop}_{\pi^{v\sigma}_t}}{f} = - \frac{1}{16(4\pi f)^2} 
\sum_B g_B \left( 2 \ell(\pi^{v\sigma}_B) + \ell(K^{v\sigma}_B) \right) 
\label{eq:decay-vs-pion_fin} \\
+\ &\frac{1}{2(4\pi f)^2} \Bigg[ \sum_X  \Bigg( \delta^{v\sigma}_V \left( \frac{\delta^{v\sigma}_V}{\delta^{\sigma\sigma}_V} - \frac{\Theta^{VF}}{2} \right)
R^{S^{\sigma\sigma}}_{\pi^{vv}\eta^{\sigma\sigma}\eta^{\prime\,\sigma\sigma}} (X_V) \nonumber \\
+\ &\frac{(\delta^{v\sigma}_V)^2}{\delta^{\sigma\sigma}_V} a^2 (\Delta^{\sigma\sigma}_V - \Delta^{vv}_V) D^{S^{\sigma\sigma}}_{\pi^{vv}\eta^{\sigma\sigma}\eta^{\prime\,\sigma\sigma},\pi^{vv}} (X_V) 
\Bigg) \ell(X_V) \nonumber \\
+\ &\delta^{\sigma\sigma}_V \sum_X R^{S^{\sigma\sigma}}_{\pi^{\sigma\sigma}\eta^{\sigma\sigma}\eta^{\prime\,\sigma\sigma}} (X_V) \ell(X_V) \nonumber \\
+\ &\Big[ (\delta^{v\sigma}_V)^2/\delta^{\sigma\sigma}_V a^2 ( \Delta^{\sigma\sigma}_V - \Delta^{vv}_V ) R^{S^{\sigma\sigma}}_{\pi^{vv}\eta^{\sigma\sigma}\eta^{\prime\,\sigma\sigma}}(\pi^{vv}_V) \nonumber \\
+\ &\left( \delta^{vv}_V - ( \delta^{v\sigma}_V )^2/ \delta^{\sigma\sigma}_V \right)
\Big] \tilde{\ell}(\pi^{vv}_V)
\Bigg] + (V\rightarrow A) \nonumber \\
+\ &\frac{1}{6(4\pi f)^2} \Bigg[ \sum_X \bigg(
- R^{S^{\sigma\sigma}}_{\pi^{vv}\eta^{\sigma\sigma}} (X_I) \nonumber \\
+\ & a^2 (\Delta^{\sigma\sigma}_I - \Delta^{vv}_I) D^{S^{\sigma\sigma}}_{\pi^{vv}\eta^{\sigma\sigma},\pi^{vv}} (X_I) 
\bigg) \ell(X_I) \nonumber \\
+\ & \sum_X R^{S^{\sigma\sigma}}_{\pi^{\sigma\sigma}\eta^{\sigma\sigma}} (X_I) \ell(X_I) \nonumber \\
+\ & a^2 ( \Delta^{\sigma\sigma}_I - \Delta^{vv}_I ) R^{S^{\sigma\sigma}}_{\pi^{vv}\eta^{\sigma\sigma}}(\pi^{vv}_I)
\tilde{\ell}(\pi^{vv}_I)
\Bigg] \,. \nonumber
\end{align}
\jab{As for the masses, we observe that double poles do not completely cancel
in the loop integrals, and the chiral behavior differs nontrivially from the
behavior in the ordinary, unmixed theory.  
The associated chiral logarithms and residues are
multiplied by combinations of LECs that vanish when valence and sea quark
actions are the same.}
%
\section{\label{sec:sum}Conclusion}
In mixed-action SChPT, we have calculated the NLO loop corrections to the
masses and decay constants of pions and kaons in all taste irreps.  
We have cross-checked all results by performing two independent calculations
and verifying the results reduce correctly when valence and sea quark actions
are the same.  
\jab{Each quantity was calculated by each of two authors, working individually.
The results were compared, and the calculations were corrected individually by
each responsible author.}
\jab{In addition, the method we use simplifies the calculations, by avoiding
the task of explicitly enumerating the vertices.  This method is explained in
Appendix C of Ref.~\cite{SWME:2011aa}.}

In the valence-valence sector, the taste pseudoscalars are Goldstone bosons in
the chiral limit, at non-zero lattice spacing, as in ordinary, unmixed SChPT.  
The NLO analytic corrections arise from tree-level contributions of the (NLO)
Gasser-Leutwyler and generalized Sharpe-Van de Water Lagrangians.  They have
the same form as in the unmixed case, with independent LECs in the
valence-valence, sea-sea, and valence-sea sectors.
The NLO loop corrections to the self-energies of the valence-valence pions and
kaons are given in Eq.~\eqref{eq:Sigvv}; those for the valence-sea pions and
kaons are given in Eqs.~\eqref{eq:Sigvs-con} and \eqref{eq:Sigvs-disc}; and
those for the decay constants are given in Eq.~\eqref{eq:decayfin}.  
\jab{Taking the same action for valence and sea quarks, these results
straightforwardly reduce to those of the ordinary, unmixed theory.  They are
also useful for deriving results in various cases of interest.}
As given \org{above}\jab{in Eqs.~\eqref{eq:Sigvv}, \eqref{eq:decayfin}}, the
results for the decay constants and valence-valence masses have the same form
as the results in ordinary, unmixed SChPT; \jab{they differ from the results of
the unmixed theory in the values of the LECs and the definitions of the
disconnected propagators, which contain terms like those in quenched (or
partially quenched) theories.}
\jab{These lead to additional terms in the final results, exemplified by
Eqs.~\eqref{eq:Sigvv_final}, \eqref{eq:decay-vv-pion_fin}, and
\eqref{eq:decay-vs-pion_fin}, for the pions of a 2+1 flavor theory.  The
corresponding chiral logarithms are of the same kind as those entering for
quenched (and partially quenched) theories; they arise from double poles in the
loop integrals.  However, no new loop integrals enter the calculations for the
mixed-action theory; the techniques developed for unmixed, partially quenched
theories are sufficient to write down the final results for various cases of
interest.}
The results in Eqs.~\eqref{eq:Sigvs-con} and \eqref{eq:Sigvs-disc}, for the
valence-sea masses, have additional corrections that vanish in the ordinary,
unmixed case\org{; the loop integrals are the same as those in the unmixed
theory}.  \jab{These are expected to be small, and analyses in the literature
to date have been performed by neglecting them.}  
\jab{The corresponding final results for the pions of a 2+1 flavor theory are
given in Eq.~\eqref{eq:Sigvs_final}.}

\jab{To summarize, the results for the mixed-action case are similar to those
for the unmixed case, and in principle no new challenges arise in using these
results in data analyses.  In practice, the utility of these results arises
from the advantages to be gained by using different species of improved
staggered fermions for the valence and sea quarks.}
\jab{For example, one could use a more highly-improved, computationally more
expensive, action for the valence quarks, to attack systematic errors due to
light-quark and gluon discretization effects, while at the same time attacking
statistical errors by using a less computationally expensive formulation for
the sea quarks, to include the effects of vacuum polarization.}
\jab{Our results explicitly parametrize the discretization effects of valence
and sea actions, and can be used to assess the advantages of mixed-action
calculations.  In closing we remark also that the large valence sector of the
staggered formulation of lattice QCD has yet to be exploited to increase
statistics on existing gauge field ensembles.}

\acknowledgements
We thank Claude Bernard for sharing his unpublished notes on
mixed-action staggered chiral perturbation theory.
\jab{We also thank the referee for helpful comments and suggestions.}
The research of W.~Lee was supported by the Creative Research
Initiatives Program (No.~20160004939) of the National Research
Foundation of Korea (NRF) funded by the Korean government (MEST).
J.A.B. is supported by the Basic Science Research Program of the
National Research Foundation of Korea (NRF) funded by the Ministry of
Education (No.~2015024974).
%
%
\section*{Appendix}
Here we present a derivation of the taste-singlet potential in Eq.~\eqref{eq:pot}.  The analysis is the same as for the ordinary, unmixed case, except that the spurion fields carry factors of the projection operators $P_{v,\sigma}$.

Consider the bilinears in Eq.~\eqref{eq:genop}.  Noting that the staggered U(1)$_{\epsilon}$ symmetry implies that $\{\gamma_s\otimes\xi_t , \gamma_5\otimes\xi_5\} = 0$, we see that taste-singlet bilinears, for which $\xi_t=\xi_I=I$, must have vector or axial spin structure, $\gamma_s = \gamma_{\mu},i\gamma_{\mu}\gamma_5$.  The taste structure of the associated four-fermion operators may be written~\cite{Lee:1999zxa}
\begin{align}
\pm \sum_{\mu}\left[\bar{\psi}_R(\gamma_{\mu}\otimes F_R)\psi_R \pm \bar{\psi}_L(\gamma_{\mu}\otimes F_L)\psi_L\right]^2\,,
\end{align}
where the positive (negative) signs apply for vector (axial) spin, and the spurion fields $F_X\rightarrow X F_X X^\dagger$ for $X=L,R\in$ SU(3) ensure that the operators are invariant under SU(3)$_L\times$SU(3)$_R$ transformations.

Enumerating all chiral singlets that are quadratic in the spurions and invariant under parity, there exists only a single nontrivial operator~\cite{Lee:1999zxa},
\begin{align}
\mr{Tr}(F_L\Sigma F_R\Sigma^\dagger)\,.
\end{align}
For the unmixed theory, setting $F_L=F_R=I$ for the taste singlet operators yields only a trivial operator.  But in the mixed case, we have $F_{L,R}=P_{v,\sigma}I$, and there are four nontrivial operators invariant under the chiral symmetry~\cite{CB:notes}:
\begin{align}
\mr{Tr}(P_v\Sigma P_v \Sigma^\dagger),\ &\mr{Tr}(P_v\Sigma P_{\sigma}\Sigma^\dagger),\ \\
\mr{Tr}(P_\sigma\Sigma P_v\Sigma^\dagger),\ &\mr{Tr}(P_{\sigma}\Sigma P_{\sigma}\Sigma^\dagger)\,.\nonumber
\end{align}
Introducing LECs, adding the results, and demanding parity invariance gives~\cite{CB:notes}
\begin{align}
C^{vv}_0 \, \mr{Tr}(P_v\Sigma P_v\Sigma^\dagger) + C^{\sigma\sigma}_0\,\mr{Tr}(P_{\sigma}\Sigma P_{\sigma}\Sigma^\dagger)\\
 +\ C^{v\sigma}_0\left[\mr{Tr}(P_v\Sigma P_{\sigma}\Sigma^\dagger) + \mr{Tr}(P_{\sigma}\Sigma P_v\Sigma^\dagger)\right]\,,\nonumber
\end{align}
where the equality of the coefficients of the last two operators follows from parity.  

Noting $P_v + P_{\sigma} = 1$ (the identity in flavor space), defining $\tau_3 = P_\sigma - P_v$, eliminating $P_{v,\sigma}$ in favor of $\tau_3$ and $1$, and collecting nontrivial operators, we have
\begin{align}
C_\mr{mix}\,\mr{Tr}(\tau_3\Sigma\tau_3\Sigma^\dagger)\,,
\end{align}
where $C_\mr{mix}\equiv \tfrac{1}{4}(C^{vv}_0 + C^{\sigma\sigma}_0 - 2 C^{v\sigma}_0)$.  In the unmixed case, $C^{vv}_0 = C^{\sigma\sigma}_0 = C^{v\sigma}_0$, $C_\mr{mix}=0$, and we recover the correct (trivial) result.
\bibliography{ref} 
\end{document}